%% file: ifacconf.tex
\documentclass{ifacconf}

\usepackage{graphicx}     
\usepackage{natbib}       
\usepackage{amsfonts}     
\usepackage{amsmath}      
\usepackage{xcolor}       
\usepackage{amssymb}      
\usepackage{enumitem}     
\usepackage{tikz,pgfplots}
\usepackage{caption}     
\usetikzlibrary{shapes.geometric,positioning,fit,plotmarks,calc,arrows}
\usepackage{circuitikz}
\usepackage{algorithm}
\usepackage{algpseudocode}

\newcommand{\col}{\operatorname{col}}
\newcommand{\diag}{\operatorname{diag}}

\begin{document}
\begin{frontmatter}

\title{Scalable Nonlinear DeePC: Bridging Direct and Indirect Methods and Basis Reduction} 


\author[TUe]{Thomas O. de Jong}
\author[TUe]{Mircea Lazar}
\author[TUe]{Siep Weiland}
\author[ETH]{Florian D\"orfler}

\address[TUe]{Eindhoven University of Technology,
Electrical Engineering Department, Eindhoven, The Netherlands 
(e-mail: {t.o.d.jong, m.lazar, s.weiland}@tue.nl).}

\address[ETH]{ETH Z\"urich,
Department of Information Technology and Electrical Engineering (ITET),
Automatic Control Laboratory (IfA), Z\"urich, Switzerland 
(e-mail: doerfler@control.ee.ethz.ch).}


\begin{abstract}                
This paper studies regularized data-enabled predictive control (DeePC) within a nonlinear framework and its relationship to subspace predictive control (SPC). The $\Pi$-regularization is extended to general basis functions and it is shown that, under suitable conditions, the resulting basis functions DeePC formulation constitutes a relaxation of basis functions SPC. To improve scalability, we introduce an SVD-based dimensionality reduction that preserves the equivalence with SPC, and we derive a reduced $\Pi$-regularization. A LASSO-based sparse basis selection method is proposed to obtain a reduced basis from lifted data. Simulations on a nonlinear van der Pol oscillator model indicate that, in the absence of noise, DeePC and SPC yield equivalent absolute mean tracking errors (AMEs) when large penalties are applied. In contrast, under noisy measurements, careful tuning of the DeePC regularization results in a reduced AME, outperforming SPC.
\end{abstract}

\end{frontmatter}
\section{Introduction}
Willems' fundamental lemma provides the theoretical foundation for data-driven control by enabling the prediction of system behavior directly from input-output data, without requiring an explicit model \citep{WILLEMS2005}. This principle underpins data-enabled predictive control (DeePC), which leverages historical trajectories to forecast future system behavior \citep{CoulsonDeePC2019}. For linear time-invariant (LTI) systems with noise-free data, equivalences between Model Predictive Control (MPC), Subspace Predictive Control (SPC), and DeePC have been established \citep{FiedlerrelationshipDeePC_SPC2021}. In practice, however, measurement noise and disturbances often corrupt collected data, breaking these equivalences and causing naive DeePC implementations to yield inaccurate predictions. Robust or regularized DeePC formulations are therefore essential.

Several strategies have been proposed to improve robustness of DeePC. Quadratic regularization has been theoretically motivated to achieve optimal and robust solutions under disturbances \citep{HUANG2021192}, while 1-norm regularization has been introduced to mitigate measurement noise \citep{CoulsonDeePC2019}. More recently, \citep{dorfler_SPC_relax} introduced the $\Pi$-regularization, a projection-based regularizer that shows the resulting regularized DeePC problem constitutes a relaxation of SPC and can achieve equivalence. This property is particularly appealing, as prior works such as \citep{FiedlerrelationshipDeePC_SPC2021} have shown that SPC often outperforms DeePC.

On the other hand, the results from \citep{dorfler_SPC_relax} suggest that indirect methods such as SPC are preferable when variance-type errors dominate, whereas direct methods like DeePC perform better under bias-type errors, which frequently arise when dealing with nonlinear systems. Extending DeePC to nonlinear dynamics has therefore become an active research area, see, for example, \citep{Markovsky_survey, VERHEIJEN2023Handbook, Bilgic, SelectDeePC2025}. Relevant to the approach of this paper, several approaches have pursued kernel based behavioral representations \citep{Molodchyk, Huang2024RoKer, Jong2025AKO} and more generally, basis functions based representations \citep{LazarBasis2024, LAZAR2024}. Simulations have shown that the corresponding nonlinear DeePC can outperform nonlinear SPC also under noisy data. However, a rigorous theoretical characterization of their relationship within this framework is still lacking.

A key challenge in the basis function framework is that nonlinear DeePC results in large-scale, nonconvex optimization problems. 
Hence, reducing the problem size without loosing performance is a crucial aspect. 
In the LTI setting, \citep{Zhang2023ReductionDeePC} employed singular value decomposition (SVD) to compress the data matrices without compromising performance. Adopting similar strategies could significantly benefit nonlinear basis function DeePC, although this has not yet been explored. Another key challenge is sparse basis function selection that accurately captures the system dynamics. Especially in kernel-based DeePC methods \citep{Huang2024RoKer, Jong2025AKO}, 
the number of basis functions grows linearly with the number of data samples, limiting the practical dataset size and making real-time or online implementation challenging, even for relatively small datasets. 

Despite the flexibility of a basis function representation, scalability to large datasets and a rigorous theoretical comparison between direct and indirect formulations remain open problems in data-driven predictive control. Motivated by these problems, this paper provides the following contributions:
\begin{itemize}
\item[C1] We extend the $\Pi$-regularization from \citep{dorfler_SPC_relax} to basis function DeePC and rigorously study the SPC--DeePC relation in this framework. Our main result shows that, in this framework, DeePC is a relaxation of SPC.
\item[C2] Inspired by \citep{Zhang2023ReductionDeePC}, we apply SVD to reduce the number of optimization variables, yielding a computationally efficient basis function DeePC formulation. We prove that the reduced basis functions DeePC problem constitutes a relaxation of the non-reduced basis function SPC problem.
\item[C3] We employ LASSO to select a reduced set of basis functions from lifted data, which can then be used in both basis functions DeePC or SPC formulations.
\end{itemize}


\subsection{Notations and basic definitions}
Let $\mathbb{R}$ and $\mathbb{N}$ denote the field of real and natural numbers. For every $c\in\mathbb{R}$ and $\mathbb{S}\subseteq\mathbb{R}$ define $\mathbb{S}_{\geq c}:= \{ k\in\mathbb{S} \mid k\geq c\}$. 
For any finite number $q\in\mathbb{N}_{\geq1}$ of column vectors or functions $\{\xi_1,\dots,\xi_q\}$ we use the operator $\text{col}(\xi_1,\dots,\xi_q):=\begin{bmatrix}
\xi_1^\top & \dots & \xi_q^\top    
\end{bmatrix}^\top$. For any vector $v \in\mathbb{R}^q$, we use $v_i$ to denote the $i$-th element of $v$ and $\|v\|_2 := \sqrt{\sum_{i=1}^q v_i^2}$. For any matrix $A\in \mathbb{R}^{n\times m}$, \(A_{i,j}\) denotes the element of \(A\) in the \(i\)-th row and \(j\)-th column, \(A_{i,:}\) denotes the \(i\)-th row of \(A\), and \(A_{:,j}\) denotes the \(j\)-th column of \(A\). For any positive $q\in\mathbb{N}$, $\col(\xi_1,\dots,\xi_q) = \begin{bmatrix}
    \xi^\top_1&\dots&\xi^\top_q
\end{bmatrix}^\top$ is a vector of stacked objects $\xi_1,\dots,\xi_q$, $\diag(\xi_1,\dots,\xi_q)$ is a diagonal matrix with $\xi_i$ on its diagonal. The identity matrix of dimension $n$ is denoted by $I_n$ and $\mathbf{0}_{n\times m}$ denotes a matrix of zeros with $n$ rows and $m$ columns. 
$\|A\|_F = \sqrt{\sum_{i,j} A_{i,j}^2}$. We write $\operatorname{Im}(A)$ and $\operatorname{row}(A)$ for the column and row spaces.


\section{Basis functions DeePC preliminaries}\label{sec:preliminaries}
We consider nonlinear discrete--time MIMO systems
\begin{subequations}\label{eqn:system}
\begin{align}
    x_{k+1} &= f(x_k,u_k), \\
    y_k &= h(x_k).
\end{align}
\end{subequations}
in state-space form with inputs $u_k\in\mathbb{R}^m$, states $x_k\in\mathbb{R}^n$ and outputs $y_k\in\mathbb{R}^p$ at time instant $k\in\mathbb{N}$. The functions $f:\mathbb{R}^n\times\mathbb{R}^m\rightarrow\mathbb{R}^n$ and $h:\mathbb{R}^n\rightarrow \mathbb{R}^p$ are assumed to be unknown. We denote by $\mathbb{X}\subseteq \mathbb{R}^n$, $\mathbb{U}\subseteq \mathbb{R}^m$ and $\mathbb{Y}\subseteq \mathbb{R}^p$ the constraint admissible sets of states, inputs, and outputs, respectively, we assume that these sets are closed.

In model predictive control (MPC), at each time instant $k$, the system model is used to compute a  sequence of predicted outputs. In \emph{indirect} data-driven predictive control, the predictor takes the form
\begin{equation}\label{eq:NARX-general}
    \mathbf{y}_{[1,N]}(k) = 
    \mathbb{F}\left(\mathbf{x}_{\mathrm{ini}}(k), 
    \mathbf{u}_{[0,N-1]}(k)\right),
\end{equation}
where $\mathbf{x}_{\text{ini}}(k) := \col \left(
    \mathbf{u}_{\mathrm{ini}}(k), 
    \mathbf{y}_{\mathrm{ini}}(k)\right)$ is a vector that contains a past window if inputs and outputs, i.e., 
\begin{align*}
    \mathbf{u}_{\mathrm{ini}}(k) &:= 
\operatorname{col}\big(u(k - T_{\mathrm{ini}}+1), \ldots, u(k - 1)\big)
\in \mathbb{R}^{(T_{\mathrm{ini}}-1) m}, \\
\mathbf{y}_{\mathrm{ini}}(k) &:= 
\operatorname{col}\big(y(k - T_{\mathrm{ini}} + 1), \ldots, y(k)\big)
\in \mathbb{R}^{T_{\mathrm{ini}} p},
\end{align*}
and $T_{\mathrm{ini}} \in \mathbb{N}_{\ge 1}$ denotes the length of the window. 
The predicted input and output sequences are defined as
\begin{align}
    \mathbf{y}_{[1,N]}(k) &:= \col(y_{1|k},\dots y_{N|k}),  \\
     \mathbf{u}_{[0,N-1]}(k) &:= \col(u_{0|k},\dots, u_{N-1|k}),
\end{align}
where for a variable $a$ the notation $a_{i|k}$ is the predicted value of $a_{k+i}$ at time instant $k$. 

Next, we summarize the basis functions framework for data-driven predictive control, introduced in \citep{LazarBasis2024}, which provides a systematic approach to constructing nonlinear multi-step predictors of the form \eqref{eq:NARX-general}. Consider a finite set of basis functions 
$\{\phi_0(\cdot), \phi_1(\cdot), \ldots, \phi_L(\cdot)\}$, with \(L \in \mathbb{N}_{\ge 1}\), shared across all MIMO nonlinear predictors. For each predicted output component $y_{i|k}\in\mathbb{R}^p$, the 
multi-step nonlinear predictor \eqref{eq:NARX-general} admits the parametrization
\begin{align*}
   y_{i|k}
&= \sum_{l=1}^L \theta_{i,l}\,
\phi_l\left(\mathbf{x}_{\mathrm{ini}}(k), \mathbf{u}_{[0,N-1]}(k)\right) \\
&= 
\begin{bmatrix}
    \theta_{i,1} & \cdots &\theta_{i,L}
\end{bmatrix}
\bar{\phi} \left(\mathbf{x}_{\mathrm{ini}}(k), \mathbf{u}_{[0,N-1]}(k)\right),
\end{align*}
where $\bar{\phi}(\cdot) := \col(\phi_1(\cdot),\dots,\phi_L(\cdot))\in\mathbb{R}^L$ and $\theta_{i,l}\in\mathbb{R}^p$ for all $i\in\{1,\dots,N\}$ and $l\in\{1,\dots,L\}$. Stacking all future predicted outputs for \(i\in \{1,\ldots,N\}\) yields the linear-in-the-parameters nonlinear autoregressive exogenous (NARX) predictor
\begin{equation}\label{eq:NARX-basis}
    \mathbf{y}_{[1,N]}(k) := \Theta \bar{\phi} \left(\mathbf{x}_{\mathrm{ini}}(k),  \mathbf{u}_{[0,N-1]}(k)\right),
\end{equation}
where $\Theta\in\mathbb{R}^{Np\times L}$. To identify the predictor parameters $\Theta$ and ensure accurate predictions, 
we excite the system in \eqref{eqn:system} using a persistently exciting input, see e.g. \citep{VERHEIJEN2023Handbook}, and collect the dataset
\[
\mathcal{D} = \{u_i, y_i\}_{i=0}^{T_{\mathrm{ini}} + T + N-1},
\]
where $T$ dictates the duration of the experiment. Next, for any $k \ge 0$ (starting time instant of the data vector) and 
\(j \ge 1\) (length of the data vector), define
\begin{subequations}
\begin{align}
    \bar{\mathbf{u}}(k,j) &:= \col\left(u(k), \ldots, u(k+j-1)\right),  \\
\bar{\mathbf{y}}(k,j) &:= \col\left(y(k), \ldots, y(k+j-1)\right).
\end{align}
\end{subequations}
Then we define the Hankel matrices:
\begin{align*}
    \mathbf{U}_p &:=
\begin{bmatrix}
\bar{\mathbf{u}}(0, T_{\mathrm{ini}}-1) & \cdots & \bar{\mathbf{u}}(T-1, T_{\mathrm{ini}}-1)
\end{bmatrix}, \\
\mathbf{Y}_p &:=
\begin{bmatrix}
\bar{\mathbf{y}}(0, T_{\mathrm{ini}}) & \cdots & \bar{\mathbf{y}}(T-1, T_{\mathrm{ini}})
\end{bmatrix}, \\
\mathbf{U}_f &:=
\begin{bmatrix}
\bar{\mathbf{u}}(T_{\mathrm{ini}}, N) & \cdots & \bar{\mathbf{u}}(T_{\mathrm{ini}} + T - 1, N)
\end{bmatrix}, \\
\mathbf{Y}_f &:=
\begin{bmatrix}
\bar{\mathbf{y}}(T_{\mathrm{ini}} + 1, N) & \cdots & \bar{\mathbf{y}}(T_{\mathrm{ini}} + T, N)
\end{bmatrix},
\end{align*}
where $T$ is the number of columns. Other data structures may be used, provided that each column is a trajectory of the system. Next, define the lifted data matrix
\begin{equation}\label{eq:Psi-matrix}
    \Phi := \bar{\phi}(\mathbf{U}_p, \mathbf{Y}_p, \mathbf{U}_f)
    \in \mathbb{R}^{L \times T},
\end{equation}
whose \(j\)-th column is obtained by evaluating 
\(\bar{\phi}\) at the corresponding trajectory 
$\begin{bmatrix}
    \mathbf{U}_p^\top, \mathbf{Y}_p^\top, \mathbf{U}_f^\top
\end{bmatrix}^\top_{j,:}$.  
Thus, \(\Phi\) contains the basis-function representation of all data 
trajectories (columns) available. Then, the nonlinear least-squares problem to learn the NARX predictor becomes
\begin{equation}\label{eq:LS-basis}
    \min_{\Theta} 
    \| \mathbf{Y}_f - \Theta \Phi\,\|_F^2,
\end{equation}
whose unique minimizer (assuming $\Phi$ has full row rank) is given by
\begin{equation}\label{eq:Theta-solution}
    \Theta^\star 
    = \mathbf{Y}_f \Phi^\dagger
    = \mathbf{Y}_f\,\Phi^\top\,(\Phi\Phi^\top)^{-1}.
\end{equation}
If $\Phi$ is not full row rank, the pseudoinverse still gives a minimum-norm solution, but the minimizer is not unique. 

Using this approach, \citep{LazarBasis2024} formulated a nonlinear \emph{indirect} data-driven predictive control scheme, similar to SPC, which we present first.
\begin{prob}[$\Phi$-\text{SPC}]\label{prob:spc_ker}
\begin{subequations}\label{eq:KerODeePC}
\begin{align}
\min_{\Xi_k} \; &\ell\left(\mathbf{y}_{[1,N]}(k),\mathbf{u}_{[0,N-1]}(k)\right) \\
\text{s.t. } \quad 
&\mathbf{Y}_f \Phi^{\dagger} \bar{\phi} \left(\mathbf{x}_{\mathrm{ini}}(k), \mathbf{u}_{[0,N-1]}(k) \right) = \mathbf{y}_{[1,N]}(k), \label{eq:bSPC_a}\\
&\big(\mathbf{y}_{[1,N]}(k),\, \mathbf{u}_{[0,N-1]}(k)\big) 
  \in \mathbb{Y}^N \times \mathbb{U}^N. \label{eq:bSPC_B}
\end{align}
\end{subequations}
\end{prob}
Secondly, \citep{LazarBasis2024} proposed a nonlinear \emph{direct} data-driven predictive control scheme, similar to DeePC.
\begin{prob}[$\Phi$-DeePC]\label{prob:DeePC_basis_nonreduced}
\begin{subequations}\label{eq:KerODeePC}
\begin{align}
\min_{\Xi_k,\mathbf{g}_k} \; &\ell \left(\mathbf{y}_{[1,N]}(k),\mathbf{u}_{[0,N-1]}(k)\right)
    + \ell_g(\mathbf{g}_k) \\
\text{s.t. } \quad 
&\Phi \mathbf{g}_k = \bar{\phi}\left(\mathbf{x}_{\mathrm{ini}}(k), \mathbf{u}_{[0,N-1]}(k) \right), \label{eq:KerODeePC_a}\\
&\mathbf{Y}_f \mathbf{g}_k = \mathbf{y}_{[1,N]}(k), \label{eq:KerODeePC_b}\\
&\big(\mathbf{y}_{[1,N]}(k),\, \mathbf{u}_{[0,N-1]}(k)\big) 
  \in \mathbb{Y}^N \times \mathbb{U}^N. \label{eq:KerODeePC_c}
\end{align}
\end{subequations}
\end{prob}
In both problems $\Xi_k = \{\mathbf{u}_{[0,N-1]}(k),\, \mathbf{y}_{[1,N]}(k)\}$. For a suitable reference $\mathbf{r}_{[1,N]}(k) := \col(r_{1|k},\dots r_{N|k})$, we typically use a quadratic cost, i.e., $\ell\left(\mathbf{y}_{[1,N]}(k),\mathbf{u}_{[0,N-1]}(k)\right):= (y_{N|k} - r_{N|k})^\top P (y_{N|k} - r_{N|k}) +  \sum_{i=0}^{N-1}(y_{i|k} - r_{i|k})^\top Q (y_{i|k} - r_{i|k})
+ \Delta u_{i|k}^\top R \Delta u_{i|k}$, where $\Delta u_{i|k}:= u_{i|k}- u_{i-1|k}$ and $\Delta u_{0|k}:= u_{0|k}- u_{k-1}$. We assume that $P \succ 0$, $Q \succ 0$ and $R \succ 0$. The choice of a suitable cost $\ell_g$ will follow. 

$\Phi$-DeePC faces the following challenges. Firstly, when the basis functions $\bar{\phi}$ provide a non-exact representation of the underlying system and/or data is corrupted by measurement noise, the concatenated matrix $\begin{bmatrix}\Phi^\top & \mathbf{Y}_f^\top\end{bmatrix}^\top$ is typically full row rank. As a result, its columns span $\mathbb{R}^{L+Np}$, and the optimization variables 
$\mathbf{u}_{[0,N-1]}(k)$ and $\mathbf{y}_{[1,N]}(k)$ are effectively free. 
This leads to poor closed-loop performance in the absence of a regularization cost $\ell_g$. For example, for the typical cost we presented after Problem~\ref{prob:DeePC_basis_nonreduced}, the optimizer frequently selects 
$\mathbf{u}_{[0,N-1]}(k) = \mathbf 0_{Nm\times 1}$ when $u_{k-1}=\mathbf{0}_{m\times 1}$ and $\mathbf{y}_{[1,N]}(k) = \mathbf{r}_{[1,N]}(k)$ as this minimizes the cost function. 
Secondly, computational efficiency poses a critical challenge in real-time, 
since $\Phi$-DeePC introduces additional optimization variables, compared with $\Phi$-SPC, 
that scale with the number of data samples $T$, i.e., 
$\mathbf{g}_k \in \mathbb{R}^T$, $\Phi \in \mathbb{R}^{L\times T}$ and $\mathbf{Y}_f \in \mathbb{R}^{Np\times T}$, which 
significantly limits the practical applicability of nonlinear $\Phi$-DeePC for large datasets. Complexity increases further with the number of basis functions, which can be large. Next, we develop solutions for addressing these challenges. 

\section{Bridging $\Phi$-SPC and $\Phi$-DeePC via the $\Pi$-regularization}\label{sec:regularization}
The $\Pi$-regularization introduced in \citep{dorfler_SPC_relax} defines
\begin{align}\label{eqn:Pi_regularization_LTI}
    \ell_{g}(\mathbf{g}_k) = \lambda \|(I-\Pi)\mathbf{g}_k\|_p,
\end{align}
where $\|\cdot\|_p$ is any p-norm, $\Pi:=  \begin{bmatrix}
    \mathbf{U}_p\\
    \mathbf{Y}_p\\
    \mathbf{U}_f\\
\end{bmatrix}^{\dagger}\begin{bmatrix}
    \mathbf{U}_p\\
    \mathbf{Y}_p\\
    \mathbf{U}_f\\
\end{bmatrix}$ and $(I-\Pi)$ is an orthogonal projector onto the null space of the three block constraint matrices. In \citep{dorfler_SPC_relax} it was shown that for sufficiently large $\lambda$, SPC and $\Pi$-regularized DeePC remain equivalent in the linear setting.

For $\Phi$-DeePC, an analogous $\Pi$-regularization that enforces equivalence with $\Phi$-SPC has not been explored. 
Motivated by the LTI case~\eqref{eqn:Pi_regularization_LTI}, we introduce a corresponding regularizer for the nonlinear, 
basis-functions setting:
\begin{align}\label{eqn:Pi_regularization}
    \ell_{g}(\mathbf{g}_k) = \lambda \,\|(I - \Phi^\dagger \Phi)\mathbf{g}_k\|_p .
\end{align}
We refer to the resulting formulation as $\Phi$-DeePC-$\Pi$, that is, $\Phi$-DeePC equipped with the extended 
$\Pi$-regularization. Next, we show that, analogous to the LTI case, $\Phi$-DeePC-$\Pi$ constitutes a relaxation of $\Phi$-SPC.

\begin{thm}[$\Phi$-DeePC-$\Pi$ relaxes $\Phi$-SPC]\label{theorem1}
\leavevmode
Consider the \emph{indirect} data-driven control Problem~\ref{prob:spc_ker} ($\Phi$-SPC) and the \emph{direct} data-driven control Problem~\ref{prob:DeePC_basis_nonreduced} with extended $\Pi$-regularization \eqref{eqn:Pi_regularization} ($\Phi$-DeePC-$\Pi$) parameterized by $\lambda \ge 0$. Assume that $\Phi$ has full row rank and that the cost $\ell$ is Lipschitz continuous with Lipschitz constant $\kappa$. 
Then, for any $\lambda\geq 0$, $\Phi$-DeePC-$\Pi$ is a relaxation of $\Phi$-SPC in the following sense:
\begin{enumerate}[label=(\roman*)]
    \item every feasible pair $(\mathbf{u}^{\star}_{[0,N-1]}(k)$, $\mathbf{y}^{\star}_{[1,N]}(k))$ of $\Phi$-SPC is feasible for $\Phi$-DeePC-$\Pi$;
    \item the optimal value of the $\Phi$-DeePC-$\Pi$ cost is a lower bound on that of the $\Phi$-SPC cost;
     \item the set of feasible solutions of $\Phi$-SPC and $\Phi$-DeePC-$\Pi$ coincide when $\lambda > \lambda^*=\kappa$.
\end{enumerate}
\end{thm}
\begin{pf}
Let $(\mathbf{u}^{\star}_{[0,N-1]}(k)$, $\mathbf{y}^{\star}_{[1,N]}(k))$ be any feasible solution of $\Phi$-SPC.  
By definition of the SPC map, it holds that
$$
\mathbf{y}_k^\star
= \mathbf{Y}_f\Phi^\dagger \bar{\phi}\left(\mathbf{x}_{\mathrm{ini}}(k) , \mathbf{u}^{\star}_{[0,N-1]}(k)\right).
$$
This is precisely the least-norm solution of
\begin{align*}
    \{\mathbf{g}_k^\star,\mathbf{y}_k^\star\}
&= \arg\min_{\mathbf{g}_k,\mathbf{y}_k}\|\mathbf{g}_k\|_2 \\
\text{s.t. } \quad 
&\Phi\mathbf{g}_k
= \bar{\phi}\left(\mathbf{x}_{\mathrm{ini}}(k), \mathbf{u}^{\star}_{[0,N-1]}(k) \right), \\ 
& \mathbf{Y}_f\mathbf{g}_k = \mathbf{y}_k,
\end{align*}
under the assumption that $\Phi$ has full row rank, so that a right inverse 
$\Phi^\dagger$ exists with $\Phi \Phi^\dagger = I$, which is a sufficient condition 
for the feasibility of the problem above. Observe that all constraint-admissible values of $\mathbf{g}_k$ in the problem above are given by 
\begin{equation}
\label{eq:decomposition}
\mathbf{g}_k
=
\underbrace{
\Phi^\dagger \bar{\phi}\left( \mathbf{x}_{\mathrm{ini}}(k) , \mathbf{u}^{\star}_{[0,N-1]}(k) \right)
}_{\mathbf{g}_k^\star}
+
(I-\Phi^\dagger\Phi)\hat{\mathbf{g}}_k,
\end{equation}
where $\hat{\mathbf{g}}_k$ is arbitrary since $(I-\Phi^\dagger\Phi)$ is an orthogonal projector onto the null space of $\Phi$. Thus, $\mathbf{g}_k=\mathbf{g}_k^\star$ if and only if $(I-\Phi^\dagger\Phi)\hat{\mathbf{g}}_k=\mathbf{0}$ which in turn is equivalent to $\|(I-\Phi^\dagger\Phi)\hat{\mathbf{g}}_k\|_p=0$ for any $p \in [1,\infty]$, since all $p$-norms vanish if and only if their argument is a zero vector. Using this equivalence, $\Phi$-SPC can be rewritten as follows:
\begin{prob}\label{prob:DeePC_basis_penalized}\begin{subequations}
\label{eq:SPC_equiv}
\begin{align}
\min_{\Xi_k} \; &\ell \left(\mathbf{y}_{[1,N]}(k),\mathbf{u}_{[0,N-1]}(k)\right) \\
\text{s.t. } \quad 
&\Phi \mathbf{g}_k = \bar{\phi}\left(\mathbf{x}_{\mathrm{ini}}(k),  \mathbf{u}_{[0,N-1]}(k)\right), \\
&\mathbf{Y}_f \mathbf{g}_k = \mathbf{y}_{[1,N]}(k), \\
&\big(\mathbf{y}_{[1,N]}(k),\, \mathbf{u}_{[0,N-1]}(k)\big) 
  \in \mathbb{Y}^N \times \mathbb{U}^N, \\
&\|(I-\Phi^\dagger\Phi)\mathbf{g}_k\|_p = 0. \label{eqn:equality_const}
\end{align}
\end{subequations}
\end{prob}
Next, we leverage Proposition~A.3 in the Appendix of~\citep{dorfler_SPC_relax}. In our setting, this implies that if the stage cost $\ell$ is Lipschitz continuous with constant $\kappa$, then for any penalty parameter $\lambda > \lambda^{\star} := \kappa$, every local minimizer of $\Phi$-DeePC-$\Pi$ is also a local minimizer of the constrained Problem~\ref{prob:DeePC_basis_penalized} and thus of $\Phi$-SPC. Consequently, replacing the hard constraint $\|(I-\Phi^\dagger\Phi)\mathbf{g}_k\|_p = 0$ with its penalized counterpart $\lambda \|(I-\Phi^\dagger\Phi)\mathbf{g}_k\|_p$ preserves equivalence with $\Phi$-SPC whenever $\lambda > \lambda^\star$. 

Finally, consider choosing $\lambda \leq \lambda^\star$. Then the penalization of the null space becomes weak, i.e., we allow $\|(I-\Phi^\dagger\Phi)\mathbf{g}_k\|_p > 0$. Therefore, the feasible set of $\Phi$-DeePC-$\Pi$ strictly contains the feasible set of Problem~\ref{prob:DeePC_basis_penalized} and thus of $\Phi$-SPC.  
This proves: ($i$) every feasible solution for $\Phi$-DeePC-$\Pi$ is feasible for $\Phi$-SPC; ($ii$) The feasible set is enlarged when $\lambda\le\lambda^{\star}$ and equivalent when $\lambda>\lambda^{\star}$ and the SPC optimal solution corresponds to a feasible $\mathbf{g}_k$ that satisfies \eqref{eqn:equality_const}. Hence, the optimal value of $\Phi$-DeePC-$\Pi$ lower bounds that of $\Phi$-SPC and this also proves ($iii$). \hfill $\blacksquare$
\end{pf}

\begin{rm}
    Theorem~\ref{theorem1} adapts Theorem~IV.6 (SPC relaxation) from \citep{dorfler_SPC_relax} to the basis function framework considered in this work. Unlike the original setting, the convexity assumptions do not carry over, since both $\Phi$-SPC and $\Phi$-DeePC are generally nonconvex for arbitrary choices of basis functions. A further difference is that we do not need to perform a convex relaxation by dropping rank or block-triangularity constraints for $\Theta$. 
\end{rm}


\section{Basis Reduction and further bridging}\label{sec:reduction}

Inspired by \citep{Zhang2023ReductionDeePC} we will utilize the properties of SVD to arrive a computationally tractable $\Phi$-DeePC reformulation.  
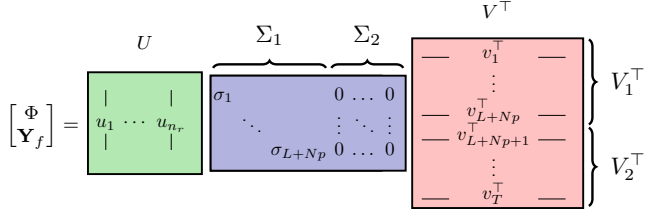
\begin{figure}[t!]
    \centering
    \input{svd_structure}
    \caption{Illustration of the singular value decomposition of data matrices assuming $T\gg Np+L$.}
    \label{fig:svd}
\end{figure}
We assume that the number of data samples satisfies $T \gg Np + L$. This ensures that the SVD of the stacked data matrix exhibits the structure
\begin{align}\label{eqn:svd_struct_econ}
    \begin{bmatrix}
    \Phi \\
    \mathbf{Y}_f 
\end{bmatrix} = U \begin{bmatrix}
    \Sigma_1 & \Sigma_2
\end{bmatrix}\begin{bmatrix}
    V_1^\top \\ V_2^\top
\end{bmatrix} = U \Sigma_1 V_1^\top,
\end{align}
where $\Sigma_1 = \diag(\sigma_1,\dots,\sigma_{L+Np})$ contains the singular values and $\Sigma_2 = \mathbf{0}_{Np+L \times T-Np-L}$, see also Fig.~\ref{fig:svd}. Since the stacked matrix need not be full row rank, some singular values may be zero. Next, we define
\begin{align}\label{eqn:reduced}
    \begin{bmatrix}
        \tilde{\Phi} \\
        \tilde{\mathbf{Y}}_f
    \end{bmatrix}
    :=
    \begin{bmatrix}
        \Phi \\
        \mathbf{Y}_f
    \end{bmatrix} V_1
    = U \Sigma_1 
    \in \mathbb{R}^{(L + Np) \times (L + Np)}.
\end{align}
This yields a reduced basis functions SPC problem.
\begin{prob}[$\tilde{\Phi}$-\text{SPC}]\label{prob:spc_ker_reduced}
\begin{subequations}\label{eq:KerODeePC}
\begin{align}
\min_{\Xi_k} \; &\ell\left(\mathbf{y}_{[1,N]}(k),\mathbf{u}_{[0,N-1]}(k)\right) \\
\text{s.t. } \quad 
&\tilde{\mathbf{Y}}_f \tilde{\Phi}^{\dagger} \bar{\phi} \left(\mathbf{x}_{\mathrm{ini}}(k), \mathbf{u}_{[0,N-1]}(k) \right) = \mathbf{y}_{[1,N]}(k), \label{eq:bSPC_a_r}\\
&\big(\mathbf{y}_{[1,N]}(k),\, \mathbf{u}_{[0,N-1]}(k)\big) 
  \in \mathbb{Y}^N \times \mathbb{U}^N. \label{eq:bSPC_B}
\end{align}
\end{subequations}
\end{prob}

Next, we establish equivalence of the reduced $\tilde{\Phi}$-SPC (Problem~\ref{prob:spc_ker_reduced}) and the non-reduced $\Phi$-SPC (Problem~\ref{prob:spc_ker}).

\begin{lem}[Equivalence of $\tilde{\Phi}$-SPC and $\Phi$-SPC]\label{lemma2}
\leavevmode
Let $T > Np + L$, and define the reduced matrices $\tilde{\Phi} = \Phi V_1$ and $\tilde{\mathbf{Y}}_f = \mathbf{Y}_f V_1$. Then $\tilde{\Phi}$-SPC and $\Phi$-SPC are equivalent.
\end{lem}

\begin{pf}
\leavevmode
Under the assumption that $T>Np+L$ the structure in Fig.~\ref{fig:svd}. is guaranteed, and it holds that
$$
\Phi^\dagger = V_1 V_1^\top \Phi^\dagger, \quad \tilde{\Phi}^\dagger = V_1^\top \Phi^\dagger.
$$
 These identities can easily be verified by using the four Moore–Penrose conditions \citep{Penrose_1955}. Next, we use the facts that $V_1^\top V_1 = I$ and $V_1 V_1^\top \Phi^\top = \Phi^\top$, which holds because $V_1 V_1^\top$ is the orthogonal projector onto the column space of $V_1$, and $\mathrm{im}(\Phi^\top) = \mathrm{row}(\Phi) \subseteq \mathrm{row}\left(\begin{bmatrix}
     \Phi^\top & \mathbf{Y}_f^\top
 \end{bmatrix}^\top\right) = \mathrm{im}(V_1)$. Finally, we obtain
$$
\tilde{\mathbf{Y}}_f \tilde{\Phi}^\dagger = \mathbf{Y}_f V_1 V_1^\top \Phi^\dagger = \mathbf{Y}_f \Phi^\dagger,
$$
which together with \eqref{eq:bSPC_a}  and \eqref{eq:bSPC_a_r} proves proves the equivalence of $\tilde{\Phi}$-SPC and $\Phi$-SPC. \hfill $\blacksquare$
\end{pf}
\begin{rem}\label{rem:relations_SPC}
Lemma~\ref{lemma2} established that $\tilde{\mathbf{Y}}_f \tilde{\Phi}^{\dagger} = \mathbf{Y}_f \Phi^{\dagger}$. Since both quantities are computed \emph{offline} and substituted into the respective Problem~\ref{prob:spc_ker_reduced} and Problem~\ref{prob:spc_ker}, the two problems are fully equivalent.
\end{rem}
Next, similar to Problem~\ref{prob:spc_ker_reduced}, we present the reduced basis-functions DeePC problem.
\begin{prob}[$\tilde{\Phi}$-DeePC]\label{prob:DeePC_basis_reduced}
\begin{subequations}\label{eq:KerODeePC}
\begin{align}
\min_{\Xi_k} \; &\ell \left(\mathbf{y}_{[1,N]}(k),\mathbf{u}_{[0,N-1]}(k)\right)
    + \ell_{\tilde{g}}(\tilde{\mathbf{g}}_k) \\
\text{s.t. } \quad 
&\tilde{\Phi} \tilde{\mathbf{g}}_k = \bar{\phi}\left(\mathbf{x}_{\mathrm{ini}}(k), \mathbf{u}_{[0,N-1]}(k) \right), \label{eq:KerODeePC_r_a}\\
&\tilde{\mathbf{Y}}_f \tilde{\mathbf{g}}_k = \mathbf{y}_{[1,N]}(k), \label{eq:KerODeePC_r_b}\\
&\big(\mathbf{y}_{[1,N]}(k),\, \mathbf{u}_{[0,N-1]}(k)\big) 
  \in \mathbb{Y}^N \times \mathbb{U}^N. \label{eq:KerODeePC_r_c}
\end{align}
\end{subequations}
\end{prob}

We now turn to establishing a relationship between the reduced formulation, $\tilde{\Phi}$-DeePC (Problem~\ref{prob:DeePC_basis_reduced}), and the non-reduced formulation, $\Phi$-DeePC (Problem~\ref{prob:DeePC_basis_nonreduced}).

\begin{lem}[The relation between $\tilde{\Phi}$-DeePC and $\Phi$-DeePC]\label{lemma3}
    Let $T > Np + L$, and define the reduced matrices $\tilde{\Phi} = \Phi V_1$ and $\tilde{\mathbf{Y}}_f = \mathbf{Y}_f V_1$. Then $\tilde{\Phi}$-DeePC and $\Phi$-DeePC have equivalent feasible sets for $\mathbf{u}_{[0,N-1]}(k)$ and $\mathbf{y}_{[1,N]}(k)$, i.e., 
    \begin{enumerate}
        \item[(i)] any feasible solution pair $(\mathbf{u}^{\star}_{[0,N-1]}(k), \mathbf{y}^{\star}_{[1,N]}(k))$ of 
        $\tilde{\Phi}$-DeePC is also feasible for $\Phi$-DeePC;
        \item[(ii)] any feasible solution pair $(\mathbf{u}^{\star}_{[0,N-1]}(k), \mathbf{y}^{\star}_{[1,N]}(k))$ of $\Phi$-DeePC is also feasible for $\tilde{\Phi}$-DeePC.
    \end{enumerate}    
\end{lem}
\begin{pf}
        First we prove $(i)$. Let $(\mathbf{u}^{\star}_{[0,N-1]}(k), \mathbf{y}^{\star}_{[1,N]}(k),\tilde{\mathbf{g}}_k^\star)$ be an arbitrary feasible solution of $\tilde{\Phi}$-DeePC and choose $\mathbf{g}_k = V_1\tilde{\mathbf{g}}_k^\star$ and substitute in \eqref{eq:KerODeePC_a} and \eqref{eq:KerODeePC_b}, i.e.,
     \begin{align}\label{eqn:non_reduced}
    \begin{bmatrix}
        \Phi\\
        \mathbf{Y}_f
    \end{bmatrix}V_1\tilde{\mathbf{g}}^\star_k =
    \begin{bmatrix}
        \tilde{\Phi}\\
        \tilde{\mathbf{Y}}_f 
    \end{bmatrix} \tilde{\mathbf{g}}^\star_k = 
    \begin{bmatrix}
       \bar{\phi}\left(\mathbf{x}_{\mathrm{ini}}(k),\mathbf{u}^\star_{[0,N-1]}(k) \right) \\
         \mathbf{y}^\star_{[1,N]}(k)
    \end{bmatrix},
    \end{align}
    which finalizes the proof for $(i)$.

    Next we prove $(ii)$. Let $(\mathbf{u}^{\star}_{[0,N-1]}(k), \mathbf{y}^{\star}_{[1,N]}(k), \mathbf{g}_k^\star)$ be an arbitrary feasible solution to Problem~\ref{prob:DeePC_basis_nonreduced} then choose $\tilde{\mathbf{g}}_k = V_1^\top  \mathbf{g}_k^\star$ and substitute in the left hand side of \eqref{eq:KerODeePC_r_a} and \eqref{eq:KerODeePC_r_b}, i.e.,
     \begin{align}\label{eqn:non_reduced}
    \begin{bmatrix}
        \tilde{\Phi}\\
        \tilde{\mathbf{Y}}_f
    \end{bmatrix} \tilde{\mathbf{g}}_k =
    \begin{bmatrix}
        \tilde{\Phi}\\
        \tilde{\mathbf{Y}} _f
    \end{bmatrix} V_1^\top  \mathbf{g}_k^\star.
    \end{align}
    Next recall from \eqref{eqn:reduced} that 
    $$
    \begin{bmatrix}
        \tilde{\Phi}\\
        \tilde{\mathbf{Y}}_f
    \end{bmatrix} =  \begin{bmatrix}
       \Phi\\
        \mathbf{Y}_f
    \end{bmatrix}V_1  = U \Sigma_1.
    $$
    Substituting this in  \eqref{eqn:non_reduced} and recalling \eqref{eqn:svd_struct_econ} yields 
    $$
    \begin{bmatrix}
        \tilde{\Phi}\\
        \tilde{\mathbf{Y}}_f
    \end{bmatrix} \tilde{\mathbf{g}}_k = U\Sigma_1 V_1^\top \mathbf{g}^\star_k = \begin{bmatrix}
        \Phi\\
        \mathbf{Y}_f 
    \end{bmatrix}\mathbf{g}^\star_k.
    $$
    This finalizes the proof for $(ii)$. \hfill $\blacksquare$
\end{pf}

\begin{rem}\label{rem:relations_DeePC}
    Lemma~\ref{lemma3} shows that the feasible sets of $\tilde{\Phi}$-DeePC and $\Phi$-DeePC coincide. However, this does not guarantee identical closed-loop behavior, since $\mathbf{g}_k$ and $\tilde{\mathbf{g}}_k$ differ. Determining for which regularization choices both formulations are identical remains an open question.
\end{rem}

Next we relate $\tilde{\Phi}$-DeePC-$\tilde{\Pi}$, i.e. $\tilde\Phi$-DeePC with the reduced regularizer 
\begin{align}\label{eqn:Pi_regularization_r}
    \ell_{\tilde{g}}(\tilde{\mathbf{g}}_k) = \lambda \,\|(I - \tilde{\Phi}^\dagger \tilde{\Phi})\tilde{\mathbf{g}}_k\|_p,
\end{align}
to non reduced $\Phi$-SPC.
\begin{thm}[$\tilde{\Phi}$-DeePC-$\tilde\Pi$ relaxes $\Phi$-SPC]
\leavevmode
\label{Theorem_final}
    Consider the non reduced \emph{indirect} data-driven control Problem~\ref{prob:spc_ker} ($\Phi$-SPC) and the reduced \emph{direct} data-driven control Problem~\ref{prob:DeePC_basis_reduced} with reduced extended $\tilde\Pi$-regularization \eqref{eqn:Pi_regularization_r} ($\tilde{\Phi}$-DeePC-$\tilde{\Pi}$) parameterized by $\lambda \ge 0$. Assume that $\Phi$ has full row rank and that the cost $\ell$ is Lipschitz continuous with Lipschits constant $\kappa$. Then, for any $\lambda\geq 0$, $\Phi$-DeePC-$\Pi$ is a relaxation of $\Phi$-SPC in the following sense:
\begin{enumerate}[label=(\roman*)]
    \item any feasible pair $(\mathbf{u}^{\star}_{[0,N-1]}(k)$, $\mathbf{y}^{\star}_{[1,N]}(k))$ of $\Phi$-SPC is feasible for $\tilde{\Phi}$-DeePC-$\tilde{\Pi}$;
    \item the optimal value of the $\tilde\Phi$-DeePC-$\tilde\Pi$ cost is a lower bound on that of the $\Phi$-SPC cost for all $\lambda\geq 0$;
    \item the set of feasible solutions of $\Phi$-SPC and $\tilde{\Phi}$-DeePC-$\tilde\Pi$ coincide when $\lambda> \lambda^*=\kappa$.
\end{enumerate}
\end{thm}
\begin{pf}
Note that $\mathrm{rank}(\tilde\Phi)=\mathrm{rank}(\Phi)$. Next, the proof follows directly by applying Theorem~\ref{theorem1} to $\tilde\Phi$-SPC and $\tilde\Phi$-DeePC-$\tilde\Pi$ and by the equivalence between $\tilde\Phi$-SPC and $\Phi$-SPC proven in Lemma~\ref{lemma2}. \hfill $\blacksquare$
\end{pf}

\begin{figure}[t!]
    \centering
    \input{implication_diagram}
    \caption{Relation diagram of the regularization  and reduction results presented in section~\ref{sec:regularization} and section~\ref{sec:reduction}} 
    \label{fig:implication}
\end{figure}
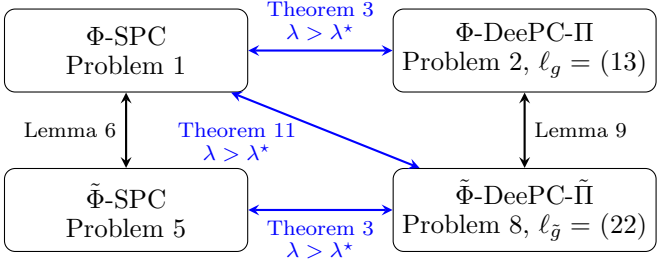

The relationships between the main results from the previous two sections are summarized in the relation diagram shown in Fig.~\ref{fig:implication}. 
We demonstrated that the developed basis reduction method does not compromise performance: \emph{the feasible set of solutions remains equivalent, and the reduced $\tilde{\Phi}$-DeePC-$\tilde{\Pi}$ formulation remains a relaxation of the original, non-reduced $\Phi$-SPC}. 

\subsection{Implications for the Linear Time-Invariant Case}
Consider the reduced linear DeePC problem with the reduced $\tilde{\Pi}$-regularization.

\begin{prob}[Reduced DeePC with $\tilde{\Pi}$-regularizer]\label{prob:DeePC_linear_reduced}
\begin{subequations}
\begin{align}
\min_{\tilde{\mathbf{g}}_k} \;& \ell \big(\mathbf{y}_{[1,N]}(k), \mathbf{u}_{[0,N-1]}(k)\big) + \|(I-\tilde{\Pi})\tilde{\mathbf{g}}_k\|_p, \\
\text{s.t. } & 
\begin{bmatrix}
\mathbf{U}_p \\
\mathbf{Y}_p \\
\mathbf{U}_f \\
\mathbf{Y}_f
\end{bmatrix} V_1 \tilde{\mathbf{g}}_k =
\begin{bmatrix}
\mathbf{u}_{\mathrm{ini}}(k) \\
\mathbf{y}_{\mathrm{ini}}(k) \\
\mathbf{u}_{[0,N-1]}(k) \\
\mathbf{y}_{[1,N]}(k)
\end{bmatrix}, \\
& (\mathbf{y}_{[1,N]}(k), \mathbf{u}_{[0,N-1]}(k)) \in \mathbb{Y}^N \times \mathbb{U}^N,
\end{align}
\end{subequations}
\end{prob}
where, the operator 
\[
\tilde{\Pi} := V_1^\top 
\begin{bmatrix} \mathbf{U}_p \\ \mathbf{Y}_p \\ \mathbf{U}_f \end{bmatrix}^\dagger
\begin{bmatrix} \mathbf{U}_p \\ \mathbf{Y}_p \\ \mathbf{U}_f \end{bmatrix} V_1,
\] 
defines the reduced $\tilde{\Pi}$-regularization. Both the decision vector $\tilde{\mathbf{g}}_k \in \mathbb{R}^{(T_{\mathrm{ini}}+N-1)m+T_\mathrm{ini}p}$ and the operator $\tilde{\Pi} \in \mathbb{R}^{((T_{\mathrm{ini}}+N-1)m+T_\mathrm{ini}p)\times ((T_{\mathrm{ini}}+N-1)m+T_\mathrm{ini}p)}$ are independent of the experiment length $T$. Lemma~\ref{lemma3} shows that the feasible set for $\mathbf{u}_{[0,N-1]}(k)$ and $\mathbf{y}_{[1,N]}(k)$ is equivalent to that of the non-reduced problem in \citep{dorfler_SPC_relax}. Finally, Theorem~\ref{Theorem_final} establishes that Problem~\ref{prob:DeePC_linear_reduced} is a relaxation of the non-reduced SPC formulation for $\lambda>\lambda^\star$.

\section{Sparse Basis Functions Selection:\\ A LASSO Kernel-Based Approach}\label{sec:kernel_approach}
Next, we leverage kernel methods and LASSO procedures to construct a reduced, data-dependent set of basis functions for $\Phi$-SPC/DeePC, where each basis function is implicitly defined by the training trajectories and the chosen kernel. To this end, consider the kernel related definitions.

\begin{defn}
A function $\mathbf{k} : \mathcal{Z} \times \mathcal{Z} \rightarrow \mathbb{R}$, where $\mathcal{Z}$ denotes the data space, is called a \emph{kernel function} if it satisfies the following properties:
\begin{enumerate}
    \item[(i)] It is symmetric, i.e., $\mathbf{k}(z_1, z_2) = \mathbf{k}(z_2, z_1)$ for all $(z_1, z_2) \in \mathcal{Z} \times \mathcal{Z}$;
    \item[(ii)] It is positive semidefinite, i.e., for any $T > 0$ and any $\{z_1, \ldots, z_T\} \subset \mathcal{Z}$, the matrix
    \[
    \mathbf{K} := 
    \begin{pmatrix}
        \mathbf{k}(z_1, z_1) & \cdots & \mathbf{k}(z_1, z_T) \\
        \vdots & \ddots & \vdots \\
        \mathbf{k}(z_T, z_1) & \cdots & \mathbf{k}(z_T, z_T)
    \end{pmatrix}
    \in \mathbb{R}^{T \times T},
    \tag{2}
    \]
    is positive semidefinite.
\end{enumerate}
\end{defn}
The matrix $\mathbf{K}$ is called the \emph{Gram matrix}. In our setting we are interested in data spaces of the form $\mathcal{Z} =\mathbb{R}^d$, where $d$ is the length of a trajectory. Two examples of kernel functions that are frequently used are the Gaussian kernel 
\begin{align}\label{eqn:kernel_gauss}
\mathbf{k}(z_i, z_j) =
\exp \left(
    -\tfrac{1}{2}
    (z_i - z_j)^{\top}
    \Sigma_{\eta}^{-1}
    (z_i - z_j)
\right),
\end{align}
where $\Sigma_\eta = \col(\eta_1,\dots,\eta_d)$ is a diagonal matrix containing hyperparameters defining the width of the kernel. Finally, we define the kernel vector, which evaluates a new data point $\mathbf{z}$ to all the data points within our dataset, i.e. 
\begin{align}
    \bar{\mathbf{k}}(\mathbf{z}) = \begin{bmatrix}
        \mathbf{k}(\mathbf{z},z_1) &
        \ldots &
        \mathbf{k}(\mathbf{z},z_T)
    \end{bmatrix}^\top.
\end{align}
\noindent
Next, consider the same setting as before, i.e., we collect a dataset $\mathcal{D} = \{u_i, y_i\}_{i=0}^{T_{\mathrm{ini}} + T + N-1}$ and construct the Hankel matrices $\mathbf{U}_p$, $\mathbf{Y}_p$, $\mathbf{U}_f$, and $\mathbf{Y}_f$. Then the Gram matrix is defined as 
\begin{equation}\label{eq:Psi-matrix}
    \mathbf{K} := \bar{\mathbf{k}}(\mathbf{U}_p, \mathbf{Y}_p, \mathbf{U}_f)
    \in \mathbb{R}^{T \times T},
\end{equation}
where the \(j\)-th column is obtained by evaluating 
\(\bar{\mathbf{k}}\) at the corresponding trajectory 
$z_j = \begin{bmatrix} \mathbf{U}_p^\top, \mathbf{Y}_p^\top, \mathbf{U}_f^\top \end{bmatrix}_{j,:}^\top$.  

Thus, kernels implicitly define a set of data-dependent basis functions. The kernel vector $\bar{\mathbf{k}}(\mathbf{z})$ can be interpreted as evaluating these basis functions at a new point $\mathbf{z}$, while the columns of the Gram matrix $\mathbf{K}$ correspond to the basis-function representation of the training trajectories. By choosing $\Phi = \mathbf{K}$ and $\bar{\phi} = \bar{\mathbf{k}}$, we can directly define the corresponding $\Phi$-SPC ($\mathbf{K}$-SPC) and $\Phi$-DeePC ($\mathbf{K}$-DeePC) problems. However, this approach yields equality constraints and optimization variables that scale with the number of data samples $T$, i.e., $\mathbf{g}_k \in \mathbb{R}^T$, $\mathbf{K} \in \mathbb{R}^{T \times T}$, and $\bar{\mathbf{k}}(\cdot) \in \mathbb{R}^T$. 

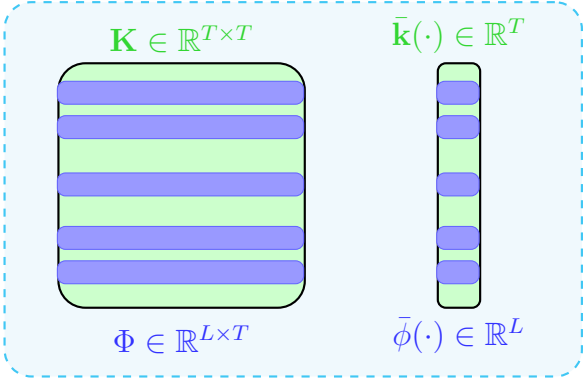
\begin{figure}[t!]
    \centering
    \input{basis_selection}
    \caption{Offline LASSO procedure for extracting a sparse basis $\bar{\phi}$ from the full kernel vector $\bar{\mathbf{k}}$, see Algorithm~\ref{alg:LASSO}.}
    \label{fig:basis_selection}
\end{figure}
\begin{figure}[t!]
    \centering
    \input{framework}
    \caption{Complete \emph{offline} framework for selecting a sparse basis and reducing redundant columns.}
    \label{fig:framework}
\end{figure}
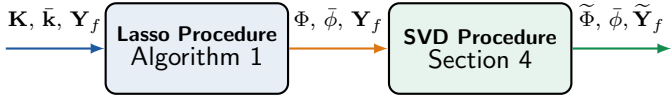
To reduce computational complexity, we construct a reduced basis from the kernel vector $\bar{\mathbf{k}}$ by identifying and removing redundant components using LASSO regression \citep{tibshirani_regression_2018}. In Fig.~\ref{fig:basis_selection}, we illustrate how LASSO is used to select basis functions from the kernel vector. This yields the next kernelized LASSO problem.

\begin{prob}[Kernelized Group LASSO]\label{prob:LASSO_ker}
\leavevmode
Given the kernel matrix $\mathbf{K}$ and target matrix $\mathbf{Y}_f$, the goal is to estimate the matrix of coefficients $\Theta$ by solving the following Group LASSO regression problem, where sparsity is enforced across the columns of $\Theta$:
\begin{align}\label{eqn:LASSO}
\Theta^\star = \arg\min_{\Theta} 
\frac{1}{2T} \|\mathbf{Y}_f - \Theta\mathbf{K} \|_F^2 
+ \alpha \sum_{j=1}^{T} \|\Theta_{:,j}\|_2,
\end{align}
where $\alpha > 0$ is a regularization parameter controlling column-wise sparsity in $\Theta$.
\end{prob}

The solution $\Theta^\star$ typically contains many zero columns. Each column of $\Theta^\star$ corresponds to one component of the lifted basis $\bar{\mathbf{k}}(z) = \col(\mathbf{k}(z,z_1), \dots, \mathbf{k}(z,z_T))$. Columns that are entirely zero indicate basis elements that do not contribute to minimizing the cost \eqref{eqn:LASSO} and can therefore be removed. Formally, let
\[
\mathcal{A} = \{j \mid \exists i, \Theta^\star_{i,j} \neq 0\}.
\]
The reduced basis is then given by
\[
\bar{\phi}(z) = \col\big( \mathbf{k}(z,z_i) : i \in \mathcal{A} \big) \in \mathbb{R}^{L},
\]
where $L:=|\mathcal{A}|$ is the reduced basis dimension, i.e. the number of columns in $\Theta^{\star}$ containing at least $1$ nonzero element. Algorithm~\ref{alg:LASSO} summarizes the procedure.

\begin{algorithm}
\caption{\emph{Offline} LASSO regression for sparse feature selection from kernelized prediction models}
\label{alg:LASSO}
\begin{algorithmic}[1]
\State \textbf{Input:} Kernel matrix $\mathbf{K} \in \mathbb{R}^{T \times T}$, target matrix $\mathbf{Y}_f \in \mathbb{R}^{pN \times T}$, regularization $\alpha > 0$, max iterations $T_{\max}$
\State Initialize $\Theta \gets \mathbf{0}_{T \times pN}$
\State Solve LASSO regression Problem~\ref{prob:LASSO_ker}, i.e., 
\For{$t = 1$ \textbf{to} $T_{\max}$}
    \State Update $\Theta$ by minimizing \eqref{eqn:LASSO}
\EndFor
\State Set $\Theta^\star \gets \Theta$
\State Identify active basis functions: $\mathcal{A} = \{j \mid \exists i, \Theta^\star_{i,j} \neq 0\}$
\State Identify number of active basis functions: $L=|\mathcal{A}|$
\State Reduced kernel vector for new input $z$:
$\bar{\phi}(z) = [k(z, z_i)]_{i \in \mathcal{A}} \in \mathbb{R}^{L}$
\State Reduced kernel data matrix:
$\Phi= [\mathbf{K}_{i,:}]_{i \in \mathcal{A}} \in \mathbb{R}^{L\times T}$
\State \textbf{Return:} $\Theta^\star$, $\mathcal{A}$, $\Phi$, $\bar{\phi}(\cdot)$ and $L$
\end{algorithmic}
\end{algorithm}
\begin{rem}\label{remark:retune}
Algorithm~\ref{alg:LASSO} directly produces the reduced data matrix $\Phi$ and feature vector $\bar{\phi}$ by selecting rows of $\mathbf{K}$ and $\bar{\mathbf{k}}$ corresponding to the LASSO-chosen basis functions. For a Gaussian kernel \eqref{eqn:kernel_gauss}, the centers $z_i$ are fixed offline, but it can be beneficial to retune the kernel widths $\Sigma_{\eta}=\diag(\eta_1,\dots,\eta_d)$ via grid search on a validation dataset. The updated feature vector is then evaluated on the dataset to construct the refined data matrix
\[
\Phi = \begin{bmatrix}
    \bar\phi(\mathbf{z}_1)& \dots & \bar\phi(\mathbf{z}_N)
\end{bmatrix},
\]
yielding a refined pair $(\bar\phi, \Phi)$. 
\end{rem}
The complete framework is illustrated in Fig.~\ref{fig:framework}. This figure summarizes the overall workflow, from constructing the initial kernel-based features to obtaining the reduced data matrices and basis function used in $\tilde\Phi$-SPC/DeePC.


\section{Numerical Example}\label{sec:example}
In this section we test the complete framework that has been proposed in this work. That 
is, starting from a kernelized model with gram matrix $\mathbf{K}$, output data Hankel matrix $\mathbf{Y}_f$ and kernel vector $\bar{\mathbf{k}}$ we use Algorithm~\ref{alg:LASSO} to get a reduced data matrix $\Phi$ and vector of basis functions $\phi$. Subsequently, we apply the SVD procedure to get $\tilde{\mathbf{Y}}_f$ and $\tilde{\Phi}$, see also Fig.~\ref{fig:framework}. As a benchmark example we use the van der Pol oscillator, a typical nonlinear benchmark. Its discrete-time dynamics are given by
\begin{subequations}\label{eqn:vanderpol}
\begin{align}
    \begin{bmatrix}
        x_1(k+1) \\
        x_2(k+1)
    \end{bmatrix}
    &=
    \begin{bmatrix}
        1 & T_s \\
        -T_s & 1 
    \end{bmatrix}
    \begin{bmatrix}
        x_1(k) \\
        x_2(k)
    \end{bmatrix}
    +
    \begin{bmatrix}
        0 \\
        T_s
    \end{bmatrix} u(k) \\
    &+
    \begin{bmatrix}
        0 \\
        T_s \mu \bigl(1 - x_1^2(k)\bigr) x_2(k)
    \end{bmatrix},  \nonumber \\
    y(k) &= \col(x_1(k)+v_1(k),x_2(k)+v_2(k)),
\end{align}
\end{subequations}
where \(u(k)\) and \(y(k)\) are the input and output at time step \(k\), We consider the following two scenarios:
\begin{enumerate}
    \item \textbf{Noise-free measurements:} \(v_1(k) = 0\) and \(v_2(k) = 0\) for all \(k\).
    \item \textbf{Noisy measurements:}  $v_1(k),\, v_2(k) \sim \mathcal{N}(0, 0.05^2)$ are independent Gaussian noise.
\end{enumerate}
Noise samples are drawn independently for each channel and time step, and the resulting noisy measurements are used in both the learning (\emph{offline}) and control (\emph{online}) phases. Throughout the experiments, we set $\mu = 1$, $T_{\mathrm{ini}}=1$ and use a sampling time $T_s = 0.1\,$s. For $\mu=1$ the system shows limit cycle behavior and is therefore a challenging example for data-driven control. In particular, we compare three setups:  
\begin{enumerate}
    \item the non-reduced basis functions SPC Problem~\ref{prob:spc_ker} ($\Phi$-SPC);
    \item reduced basis functions DeePC Problem~\ref{prob:DeePC_basis_reduced} with a simple $\lambda\|\tilde{\mathbf{g}}_k\|_2^2$ regularizer ($\tilde{\Phi}$-DeePC-$2$); 
    \item reduced basis functions DeePC Problem~\ref{prob:DeePC_basis_reduced} with the proposed regularizer $\lambda \,\|(I - \tilde{\Phi}^\dagger \tilde{\Phi})\tilde{\mathbf{g}}_k\|_2$, ($\tilde{\Phi}$-DeePC-$\tilde\Pi$).   
\end{enumerate}
For both regularizers, we compare the performance for different values of $\lambda$ in a reference tracking example. Let 
$\mathbf{y}_{\Phi\text{-SPC}}(k)$, 
$\mathbf{y}_{\tilde{\Phi}\text{-DeePC-}2}(k)$, and 
$\mathbf{y}_{\tilde{\Phi}\text{-DeePC-}\tilde{\Pi}}(k)$ 
denote the output evolutions of the system in closed loop with the three controllers. Let $T_{\text{sim}}\in\mathbb{N}$ denote the simulation duration; then the \emph{absolute mean error} (AME) of a method with output $\mathbf{y}(k)$ relative to reference $\mathbf{r}_k$ is defined as
\[
\text{AME}(\mathbf{y}) = \frac{1}{T_{\text{sim}}} \sum_{k=1}^{T_{\text{sim}}} \big| \mathbf{y}(k) - \mathbf{r}_k \big|.
\]
\noindent The \emph{absolute mean error to SPC} (AME$_{\text{spc}}$) for a method with output $\mathbf{y}(k)$ is defined as
\[
\text{AME}_{\text{spc}}(\mathbf{y}) = \frac{1}{T_{\text{sim}}} \sum_{k=1}^{T_{\text{sim}}} \big| \mathbf{y}_{\Phi_{\text{spc}}}(k) - \mathbf{y}(k) \big|.
\]

\noindent All data-driven predictive control methods are implemented in CasADi and solved with IPOPT. The numerical results were generated on a machine equipped with a 13th Gen Intel\textsuperscript{\textregistered} Core\texttrademark{} i7-1370P CPU at 1.90\,GHz.


To train the model, we excite the system \eqref{eqn:vanderpol} with a multisine input signal and collect a dataset $\mathcal{D}_{\text{train}} := \{u_i,y_i\}_{i=1}^{T}$ with $T=2000$ samples. This gives the training data matrices $\mathbf{U}^t_p$, $\mathbf{Y}^t_p$, $\mathbf{U}^t_f$ and $\mathbf{U}^t_f$. An additional validation dataset of $2000$ samples $\mathcal{D}_{\text{val}} := \{u_i,y_i\}_{i=T+1}^{2T}$ with $T=2000$ samples giving $\mathbf{U}^v_p$, $\mathbf{Y}^v_p$, $\mathbf{U}^v_f$ and $\mathbf{U}^v_f$. We used the Gaussian kernel \eqref{eqn:kernel_gauss} where $\Sigma_\eta := \diag(\eta_1,\eta_2,\dots,\eta_{n+mN}) \in \mathbb{R}^{(n+mN)\times (n+mN)}$ are tuned by performing a grid search using the validation data set $\mathcal{D}_{\text{val}}$, see Remark~\ref{remark:retune}.

To obtain a sparse set of basis functions, we apply Algorithm~\ref{alg:LASSO}. The LASSO problem~\ref{prob:LASSO_ker} is solved using \texttt{MultiTaskLASSO} from the \texttt{scikit-learn} Python library with $\alpha = 0.01$ and $T_{\text{max}} = 5000$. The singular value decomposition required in Step~2 of the framework shown in Fig.~\ref{fig:framework} is computed using \texttt{np.linalg.svd} from \texttt{NumPy}.



\begin{table*}[t!]
\centering
\captionsetup{justification=centering}
\caption{Number of basis dimensions and resulting DeePC decision variables and offline CPU times refer to the each of the reduction steps in Fig.~\ref{fig:framework}.}
\label{tab:model_cpu}

\begin{tabular}{c|cc|cc|cc|c}
\noalign{\hrule height 0.5pt}
\textbf{Procedure} &
\multicolumn{2}{c|}{\textbf{\# Basis Functions}} &
\multicolumn{2}{c|}{\textbf{\# Optimization Variables}} &
\multicolumn{2}{c|}{\textbf{Offline CPU Time}} &
\textbf{Notes} \\
& noise-free & noisy & noise-free & noisy & noise-free & noisy & \\ 
\noalign{\hrule height 0.5pt}

Initial 
& $2000$ & $2000$ 
& $2030$ & $2030$ 
& -- & --
& Initial model \\

LASSO   
& $60$ & $55$   
& $2030$ & $2030$ 
& $37.7\,$s & $52.5\,$s
& Basis selection \\

SVD     
& $60$ & $55$   
& $90$ & $80$ 
& $0.306\,$s & $0.275\,$s
& Dimensionality reduction \\

\noalign{\hrule height 0.5pt}
\end{tabular}
\end{table*}

\begin{table}[t!]
\centering
\captionsetup{justification=centering}
\caption{AME and AME$_{\text{spc}}$ for noise-free data.}
\label{tab:deeppc_convergence}
\begin{tabular}{c|cc|cc}
\noalign{\hrule height 0.5pt} 
\textbf{$\lambda$} &
\multicolumn{2}{c|}{$\tilde\Phi$-DeePC-$\tilde \Pi$} &
\multicolumn{2}{c}{$\tilde\Phi$-DeePC-$2$} \\
 & AME & AME$_{\text{spc}}$  & AME & AME$_{\text{spc}}$  \\
\noalign{\hrule height 0.5pt} 
$10^{3}$  & $1.1529$ & $0.0408$ & $1.1873$ & $0.1152$ \\
$10^{6}$  & $1.1435$ & $0.0001$ & \textcolor{red}{$\mathbf{\times}$} & \textcolor{red}{$\mathbf{\times}$} \\
$10^{9}$  & \textcolor{green!70!black}{$1.1435$} & $0.0000$ & \textcolor{red}{$\mathbf{\times}$} & \textcolor{red}{$\mathbf{\times}$} \\
\noalign{\hrule height 0.5pt} 
\end{tabular}
\end{table}

\begin{figure}[t!]
  \centering
  \includegraphics[
      width=0.99\columnwidth,
      trim=10 12 10 10,      
      clip
  ]{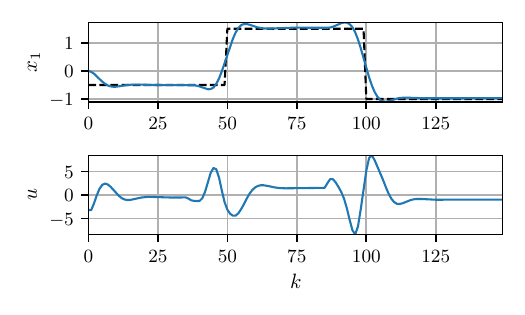}
  \caption{Tracking example for $\tilde\Phi$-DeePC-$\tilde\Pi$ with $\lambda=10^9$ for noise-free data.}
  \label{fig:mpc_results}
\end{figure}

\paragraph*{Scenario 1. Noise-free data}
Table~\ref{tab:deeppc_convergence} reports the AME and AME$_{\text{spc}}$ for both regularization schemes across different values of~$\lambda$ for the tracking problem shown in Fig.~\ref{fig:mpc_results}. The results show that the solution of the reduced and regularized $\tilde{\Phi}$-DeePC-$\tilde{\Pi}$ problem converges to the non-reduced $\Phi$-SPC solution as $\lambda$ becomes large, thereby validating Theorem~~\ref{Theorem_final}. In contrast, the standard $\lambda\|\tilde{\mathbf{g}}_k\|_2^2$ regularizer does not enforce such convergence: its deviation from the SPC trajectory increases with~$\lambda$, and the controller becomes unstable for $\lambda \ge 10^6$. The corresponding closed-loop trajectory of $\tilde{\Phi}$-DeePC-$\tilde \Pi$ with $\lambda=10^9$ is shown in Fig.~\ref{fig:mpc_results}.

\begin{table}[t!]
\centering
\captionsetup{justification=centering}
\caption{AME for noisy data.}
\label{tab:deeppc_vs_spc_noise}
\begin{tabular}{c|c|c|c}
\noalign{\hrule height 0.5pt} 
\textbf{$\lambda$} &
$\tilde\Phi$-DeePC-$\tilde\Pi$ & 
$\tilde\Phi$-DeePC-$2$ & 
$\Phi$-SPC \\
\noalign{\hrule height 0.5pt} 
$10^{3}$ & \textcolor{green!70!black}{$1.1143$} & $1.1646$ & $1.1216$   \\
$10^{6}$ & 
$1.1198$ & \textcolor{red}{$\mathbf{\times}$} & $1.1216$   \\
$10^{9}$ & $1.1277$ & \textcolor{red}{$\mathbf{\times}$} & $1.1216$   \\
\noalign{\hrule height 0.5pt} 
\end{tabular}
\end{table}

\begin{figure}[t!]
  \centering
  \includegraphics[
      width=0.99\columnwidth,
      trim=10 12 10 10,      
      clip
  ]{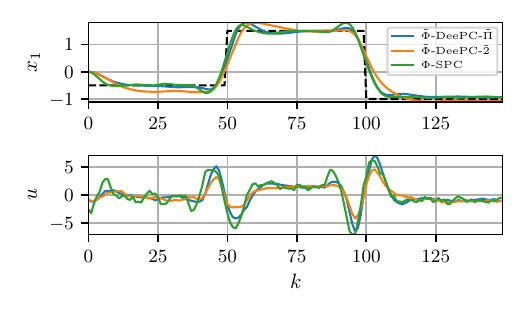}
  \caption{Tracking example for $\tilde\Phi$-DeePC-$\tilde\Pi$, $\tilde\Phi$-DeePC-$2$ with $\lambda=10^4$ and $\Phi$-SPC for noisy data.}
  \label{fig:mpc_results_noise}
\end{figure}

\paragraph*{Scenario 2. Noisy data}
Fig.~\ref{fig:mpc_results_noise} shows the same tracking experiment in the presence of measurement noise, comparing $\Phi$-SPC, $\tilde{\Phi}$-DeePC-$\tilde{\Pi}$, and $\tilde{\Phi}$-DeePC-2, which became unstable for $\lambda\geq10^6$. Table~\ref{tab:deeppc_vs_spc_noise} reports the corresponding AME values for all methods. 
Across all tested regularization levels, $\tilde{\Phi}$-DeePC-$\tilde{\Pi}$ consistently outperforms $\tilde{\Phi}$-DeePC-2. Moreover, for $\lambda \leq 10^6$, it even surpasses the performance of $\Phi$-SPC. This is contrary to the noise-free case, where $\Phi$-SPC was always superior or matched by $\tilde{\Phi}$-DeePC-$\tilde{\Pi}$ for sufficiently large~$\lambda$. 
These results also highlight a key difference from the noise-free setting: while taking $\lambda$ as large as possible was optimal there, this is no longer the case under noisy measurements. This motivates extensions of $\lambda$-tuning methods for linear DeePC \citep{Lazar_off_set} to $\Phi$-DeePC. 

Finally, the average \emph{online} CPU times for the two controllers during the tracking example are as follows: $\Phi$-SPC requires \textbf{0.0046\,s} in the noise-free case and \textbf{0.0073\,s} with noisy data, whereas $\tilde{\Phi}$-DeePC-$\tilde{\Pi}$ requires \textbf{0.0277\,s} and \textbf{0.0196\,s}, respectively. All controllers achieve a computation time lower than $T_s=0.1\,$s.
While $\Phi$-SPC remains faster overall, $\tilde{\Phi}$-DeePC-$\tilde{\Pi}$ offers advantages in handling noisy data.
Moreover, since IPOPT is a general-purpose nonlinear solver, there remains significant room to develop a tailored fast solver 
for $\tilde\Phi$-DeePC-$\tilde{\Pi}$, whose structure involves only nonlinear equality constraints.

\section{Conclusion}
This paper addressed three key challenges of nonlinear DeePC in the basis-function framework. We proposed the extended $\Pi$-regularizer, showing that the resulting regularized nonlinear DeePC problem constitutes a relaxation of nonlinear SPC. An SVD-based data reduction preserves this relaxation while decoupling problem size from dataset length, making large-scale optimization tractable. We also introduced a method to select a reduced set of basis functions via LASSO and kernel techniques, enabling efficient real-time implementation. Simulations on a nonlinear van der Pol oscillator model demonstrated that the reduced regularized DeePC can outperform non-reduced SPC under measurement noise.

Future work includes performing basis reduction efficiently online. This would tailor the basis functions to the initial condition and reference of the predictive control problem.



\bibliography{ifacconf}             
\end{document}

%% file: svd_structure.tex
\scalebox{0.9}{
\begin{tikzpicture}[font=\sffamily, every node/.style={align=center}]

\tikzset{
    block/.style={draw, thick, minimum width=1.5cm, minimum height=1.5cm, fill=#1!30},
    smallblock/.style={draw, thick, minimum width=1.5cm, minimum height=1cm, fill=#1!30},
    label/.style={font=\small\bfseries}
}

\newlength{\blockw}
\newlength{\blockh}
\setlength{\blockw}{2.5cm}
\setlength{\blockh}{1.4cm}
\newlength{\blockwV}
\setlength{\blockwV}{2.5cm}
\newlength{\blockhV}
\setlength{\blockhV}{2.5cm}

\node[block=green!70!black] (U) {
    \scalebox{0.8}[0.8]{$\begin{matrix}
        \rule{0.5pt}{0.3cm} & & \rule{0.5pt}{0.3cm} \\
        u_1 & \cdots & u_{n_r} \\
        \rule{0.5pt}{0.3cm} & & \rule{0.5pt}{0.3cm}
    \end{matrix}$}
};

\node[block=green!80!black, minimum width=\blockw, minimum height=\blockh, right=0.1cm of U] (Sigma) {};

\node[block=blue!60!black, minimum width=\blockw, minimum height=\blockh, anchor=north west] 
    at (Sigma.north west) (S1) {
    \makebox[\blockw][c]{\scalebox{0.8}[0.8]{$\begin{matrix}
        \sigma_1 & & &  0 & \dots & 0 \\
        & \ddots & &  \vdots  & \ddots & \vdots\\
        & & \sigma_{L+Np} &  0 & \dots  & 0
    \end{matrix}$}}
};

\node[block=red!80!white, minimum width=\blockwV, minimum height=\blockhV, inner sep=-0.5pt, outer sep=-0.5pt, right=0.45cm of Sigma] (Vt) {\scalebox{0.8}[0.8]{$\begin{matrix}
    \rule{0.5cm}{0.5pt}  & v^\top_1 & \rule{0.5cm}{0.5pt}   \\
    & \vdots & \\
        \rule{0.5cm}{0.5pt}  & v^\top_{L+Np} & \rule{0.5cm}{0.5pt}   \\
        \rule{0.5cm}{0.5pt}  & v^\top_{L+Np+1} & \rule{0.5cm}{0.5pt}   \\
    & \vdots & \\
        \rule{0.5cm}{0.5pt}  & v^\top_{T} & \rule{0.5cm}{0.5pt}   \\
\end{matrix}$}
};

\node[label, above=2mm of U.north] {$U$};
\node[label, above=2mm of Vt.north] {$V^\top$};
\node[label] at ([xshift=-15mm,yshift=-8mm] U.north) {$\begin{bmatrix}
    \Phi \\
    \mathbf{Y}_f
\end{bmatrix} = $};

\usetikzlibrary{decorations.pathreplacing} 

\draw[decorate,decoration={brace, amplitude=3.0pt}, line width=1pt] 
    ([xshift=1mm,yshift=0mm] Vt.north east) -- ([xshift=1mm,yshift=-1.25cm] Vt.north east) 
    node[midway,right=1.8mm] {$V_1^\top$};

\draw[decorate,decoration={brace, amplitude=3.0pt}, line width=1pt] 
    ([xshift=1mm,yshift=-1.3cm] Vt.north east) -- ([xshift=1mm,yshift=-2.45cm] Vt.north east) 
    node[midway,right=1.8mm] {$V_2^\top$};

\draw[decorate,decoration={brace, amplitude=3pt}, line width=1pt] 
    ([xshift=-11mm,yshift=1mm] S1.north east) -- ([xshift=-0.5mm,yshift=1mm] S1.north east) 
    node[midway,above=2mm] {$\Sigma_2$};

    \draw[decorate,decoration={brace, amplitude=3pt}, line width=1pt] 
    ([xshift=0.5mm,yshift=1mm] S1.north west) -- ([xshift=17mm,yshift=1mm] S1.north west) 
    node[midway,above=2mm] {$\Sigma_1$};

\end{tikzpicture}}

%% file: implication_diagram.tex
\begin{tikzpicture}[
    box/.style={
        draw,
        rounded corners,
        minimum width=3.2cm,
        minimum height=1.1cm,
        align=center
    },
    arrow/.style={>=stealth, thick},
]

\node[box] (spc) {$\Phi$-SPC\\Problem~\ref{prob:spc_ker}};
\node[box, below=1.0cm of spc] (tspc) {$\tilde{\Phi}$-SPC\\Problem~\ref{prob:spc_ker_reduced}};

\node[box, right=1.9cm of spc] (deepc) {$\Phi$-DeePC-$\Pi$\\Problem~\ref{prob:DeePC_basis_nonreduced}, $\ell_g =$ \eqref{eqn:Pi_regularization}};
\node[box, below=1.0cm of deepc] (tdeepc) {$\tilde{\Phi}$-DeePC-$\tilde{\Pi}$\\Problem~\ref{prob:DeePC_basis_reduced}, $\ell_{\tilde{g}} =$ \eqref{eqn:Pi_regularization_r}};

\draw[<->,arrow] (spc) -- (tspc) node[midway,left] {\small Lemma~\ref{lemma2}};
\draw[<->,arrow, black] (deepc) -- (tdeepc) node[midway,right, text=black] {\small \shortstack{Lemma~\ref{lemma3} }};

\draw[<->,arrow, blue] (tdeepc) -- (spc) node[midway,left, text=blue, xshift=-0.25cm, yshift=-0.15cm] {\small \shortstack{ Theorem~\ref{Theorem_final} \\ $\lambda > \lambda^\star$ }};

\draw[arrow,<->, blue] (deepc) -- (spc)
    node[midway, above, text=blue] {\small \shortstack{Theorem~\ref{theorem1} \\ $\lambda > \lambda^\star$}};

\draw[arrow,<->, blue] (tdeepc) -- (tspc)
    node[midway, below, text=blue] {\small\shortstack{Theorem~\ref{theorem1} \\ $\lambda > \lambda^\star$}};

\end{tikzpicture}

%% file: basis_selection.tex
\resizebox{0.95\linewidth}{!}{%
\begin{tikzpicture}[
    font=\large,
    mat/.style={draw, thick, rounded corners=10pt},
    strip/.style={rounded corners=3pt, fill=blue!40, draw=blue!60},
    >=latex
]

\node[mat, dashed, draw=cyan!60, fill=cyan!5,
      minimum width=7.5cm, minimum height=4.9cm,
      anchor=north west] (Lbox) at (0,0) {};

\node[anchor=south, font=\large\bfseries]
    at ($(Lbox.north)+(0,0)$)
    {Offline LASSO procedure \((\mathbf{K},\bar{\mathbf{k}}\rightarrow \Phi,\bar{\phi})\)};

\node[mat, fill=green!20,
      minimum width=3.2cm, minimum height=3.2cm,
      anchor=north west] (Kmat)
      at ($(Lbox.north west)+(0.7,-0.8)$) {};

\foreach \y in {0.55,1.0,1.75,2.45,2.9} {
    \draw[strip]
       ($(Kmat.north west)+(0,-\y)$)
       rectangle ++(3.2,0.30);
}

\node[mat, rounded corners=3pt, fill=green!20,
      minimum width=0.55cm, minimum height=3.2cm,
      anchor=north west] (kvec)
      at ($(Kmat.north east)+(1.7,0)$) {};

\foreach \y in {0.55,1.0,1.75,2.45,2.9} {
    \draw[strip]
       ($(kvec.north west)+(0,-\y)$)
       rectangle ++(0.55,0.30);
}

\node[anchor=south, text=black!20!green!80, font=\large]
    at (Kmat.north) {\(\mathbf{K}\in\mathbb{R}^{T\times T}\)};
\node[anchor=south, text=black!20!green!80, font=\large]
    at (kvec.north) {\(\bar{\mathbf{k}}(\cdot)\in\mathbb{R}^{T}\)};

\node[anchor=south, font=\large, text=blue!80, yshift=-0.7cm]
    at (Kmat.south) {\(\Phi\in\mathbb{R}^{L\times T}\)};
\node[anchor=south, font=\large, text=blue!80, yshift=-0.7cm]
    at (kvec.south) {\(\bar{\phi}(\cdot)\in\mathbb{R}^{L}\)};

\end{tikzpicture}
} 

%% file: framework.tex
\begin{tikzpicture}[>=latex, font=\sffamily, node distance=0.2cm and 1.3cm]

\definecolor{myblue}{RGB}{30,90,160}
\definecolor{mygreen}{RGB}{20,140,80}
\definecolor{myorange}{RGB}{220,120,20}

\tikzstyle{block} = [
    draw,
    thick,
    rounded corners,
    align=center,
    minimum width=2.4cm,
    minimum height=1.2cm,
    inner sep=4pt
]

\node[align=center] (inputlbl) at (-3,0.3) {};

\node[block, fill=myblue!10, right=of inputlbl] (lasso) {\small
    \textbf{Lasso Procedure}\\[-2pt]
    Algorithm~\ref{alg:LASSO}
};

\node[block, fill=mygreen!10, right=of lasso] (svd) {
    \textbf{\small SVD Procedure}\\[-2pt]
    Section~\ref{sec:reduction}
};

\node[align=center, right=of svd] (outlbl) {};

\draw[->, thick, myblue] (inputlbl.east) -- node[above, yshift=2pt] {\small \textcolor{black}{$\mathbf{K},\, \bar{\mathbf{k}},\,\mathbf{Y}_f$}} (lasso.west);
\draw[->, thick, myorange] (lasso.east) -- node[above, yshift=2pt] {\small \textcolor{black}{$\Phi,\, \bar{\phi},\,\mathbf{Y}_f$}} (svd.west);
\draw[->, thick, mygreen] (svd.east) -- node[above, yshift=2pt] {\small \textcolor{black}{$\widetilde{\Phi},\, \bar{\phi},\widetilde{\mathbf{Y}}_f$}} (outlbl.west);

\end{tikzpicture}